\documentclass[leqno,a4paper,11pt]{amsart}

\usepackage{hyperref}
\usepackage{amsthm}
\usepackage{amsmath}
\usepackage{amssymb}
\usepackage{a4wide}
\usepackage{graphics}
\usepackage{enumerate}

\newcommand{\Con}{\ensuremath{\mathcal{C}}}

\newcommand{\Cinf}{\ensuremath{\mathcal{C}^\infty}}

\newcommand{\D}{\ensuremath{{\mathcal D}}}

\newcommand{\E}{\ensuremath{{\mathcal E}}}

\newcommand{\mb}[1]{\ensuremath{\mathbb{#1}}}
\newcommand{\N}{\mb{N}}

\newcommand{\R}{\mb{R}}

\newcommand{\Z}{\mb{Z}}

\newcommand{\G}{\ensuremath{{\mathcal G}}}

\newcommand{\EM}{\ensuremath{{\mathcal E}_{\mathrm{M}}}}

\newcommand{\NN}{\ensuremath{{\mathcal N}}}

\newcommand{\lara}[1]{\langle #1 \rangle}

\renewcommand{\d}{\ensuremath{\partial}}

\newfont{\bl}{msbm10 scaled \magstep2}

\newcommand{\beq}{\begin{equation}}
\newcommand{\eeq}{\end{equation}}

\newcommand{\notmid}{\mid\kern-0.5em\not\kern0.5em}

\newcommand{\Ga}{\Gamma}

\newcommand{\eps}{\varepsilon}

\newenvironment{pr}{\begin{proof}[\textbf{Proof:}] \ }{\end{proof}}
\newtheorem{thm}{Theorem}[section]
\newtheorem{lem}[thm]{Lemma}
\newtheorem{prop}[thm]{Proposition}

\newtheorem{cor}[thm]{Corollary}

\newtheorem{defi}[thm]{Definition}

\newcommand{\dd}{{\mathrm{d}}}
\newcommand{\diam}{{\mathrm{diam}}}
\newcommand{\lip}{{\mathrm{Lip}}}
\newcommand{\Xe}{\mathfrak{X}_\eps}
\renewcommand{\d}{{\mathrm{d}}}
\newcommand{\Up}{U_+}

\newcommand{\GaG}{\Gamma_\G}
\newcommand{\ep}{\varepsilon}
\newcommand{\Rt}{\tilde{\R}}

\makeatletter
\def\subsection{\@startsection{subsection}{2}%
  \z@{.5\linespacing\@plus.7\linespacing}{.1\linespacing}%
  {\normalfont\bfseries}}
  \makeatother

\begin{document}

\title[Nonexpanding impulsive gravitational waves with $\Lambda$]{%
Geodesics in nonexpanding impulsive gravitational waves with $\Lambda$, II%
}

\author{Clemens S\"amann \and Roland Steinbauer}

\address{\textsc{Faculty of Mathematics, University of Vienna, Austria}}

\email{clemens.saemann@univie.ac.at} 
\email{roland.steinbauer@univie.ac.at}

\begin{abstract}
We investigate all geodesics in the entire class of nonexpanding impulsive 
gravitational waves propagating in an (anti-)de Sitter universe using the 
distributional metric. We extend the regularization approach of part 
I ({\it Geodesics in nonexpanding impulsive gravitational waves with ${\Lambda}$, Part I}, Class.\ Quantum Grav., 
33(11):115002, 2016) to a full nonlinear distributional analysis within the 
geometric theory of generalized functions. We prove global existence and 
uniqueness of geodesics that cross the impulsive wave and hence geodesic 
completeness in full generality for this class of low regularity spacetimes.  This, in particular, 
prepares the ground for a mathematically rigorous account  on  the 'physical 
equivalence' of the continuous with the distributional `from' of the metric.  
 \vskip 1em
  
  \medskip
  
  
\end{abstract}

\date{\today}
\maketitle


\section{Introduction}\label{sec:intro}

Impulsive gravitational waves are exact radiative spacetimes of general 
relativity that provide theoretic models of short but violent bursts of 
gravitational radiation, which are also of significant interest to quantum 
theories of gravity. Originally introduced by R.\ Penrose (e.g.\ \cite{Pen:72}) 
their construction and physical properties have been extensively studied by many 
authors, for an overview see e.g.\ \cite[Chapter 20]{GP:09}, as well as 
\cite{BH:03,P:02}. Apart from their physical significance these models are also 
interesting from a purely mathematical point of view since they are described by 
metrics of low regularity. More precisely impulsive gravitational waves have 
been described by two `forms' of the metric, one (locally Lipschitz-)continuous, 
the other one even distributional. In particular, the `physical equivalence' of 
these two descriptions has been established in several families of these models 
\emph{in a formal way}, leaving open some quite subtle issues in low regularity 
Lorentzian geometry.

In this work we are especially interested in \emph{nonexpanding} impulsive 
gravitational waves which propagate on a \emph{cosmological background} of 
constant curvature, that is on de Sitter (with cosmological constant 
$\Lambda>0$) or anti-de Sitter ($\Lambda<0$) universe. These models have come 
into focus with the pioneering work of Hotta and Tanaka \cite{HT:93}, who 
performed an ultrarelativistic boost to the Schwarzschild--(anti-)de~Sitter 
solution to obtain a nonexpanding spherical impulsive gravitational wave 
generated by a pair of null monopole particles. Since then many more such 
solutions have been found, see e.g.\ the review in \cite[Section 2]{PSSS:15}.
The corresponding \emph{continuous} form of the metric is 
given by (\cite{P:98,PG:99})
\begin{equation}\label{cont}
\d s^2= \frac{2\,|\d Z+\Up(H_{,Z\bar Z}\d Z+{H}_{,\bar Z\bar Z}\d\bar 
Z)|^2-2\,\d
U\d V}
{[\,1+\frac{1}{6}\Lambda(Z\bar Z-UV+\Up G)\,]^2}.
\end{equation}
Here $H(Z,\bar Z)$ is an arbitrary real-valued function, 
${G(Z,\bar Z)= ZH_{,Z}+\bar ZH_{,\bar Z}-H}$ and $\Up=\Up(U)=0$
for $U\leq 0$ and $\Up(U)=U$ for $U\geq0$ is the \emph{kink function}. The metric 
\eqref{cont} is most easily generated from the conformally flat form 
of the constant curvature background 
\begin{equation}\label{backgr}
\d s_0^2= \frac{2\,\d\eta\,\d\bar\eta
-2\,\d{\mathcal U}\,\d{\mathcal 
V}}{[\,1+{\frac{1}{6}}\Lambda(\eta\bar\eta-{\mathcal U}{\mathcal
 V})\,]^2}\,,
\end{equation}
where ${\mathcal U}, \mathcal{V}$ are the usual null and $\eta,\bar\eta$ the
usual complex spatial coordinates. More precisely applying the transformation
\begin{equation}\label{ro:trsf}
 {\mathcal U}=U\,,\quad 
 {\mathcal V}=
 \begin{cases}
  V &\mbox{for ${\mathcal U}<0$}\\               
  V+H+UH_{,Z}H_{,\bar Z}  &\mbox{for ${\mathcal U}>0$}       
 \end{cases}
 \,,\quad
 \eta=
 \begin{cases}
  Z &\mbox{for ${\mathcal U}<0$}\\  
  Z+UH_{,\bar Z} &\mbox{for ${\mathcal U}>0$}  
 \end{cases}
\end{equation}
to \eqref{backgr} separately for negative and positive values of ${\mathcal 
U}$ one formally obtains \eqref{cont}. For all details see 
the introduction to part I, i.e., \cite[Section 1]{SSLP:16}. The corresponding 
distributional form 
\begin{equation}
\d s^2= \frac{2\,\d\eta\,\d\bar\eta-2\,\d {\mathcal U}\,\d {\mathcal V}
+2H(\eta,\bar\eta)\,\delta({\mathcal U})\,\d {\mathcal U}^2}
{[\,1+\frac{1}{6}\Lambda(\eta\bar\eta-{\mathcal U}{\mathcal V})\,]^2}
 \label{confppimp}
\end{equation}
is \emph{formally} derived by writing \eqref{ro:trsf} in the form of a  
`discontinuous coordinate transform' using the Heaviside function $\Theta$ 
\begin{equation} \label{trans}
 {\mathcal U}=U\,,\quad
 {\mathcal V}=V+\Theta\,H+\Up\,H_{,Z}H_{,\bar Z}\,, \quad
 \eta=Z+\Up\,H_{,\bar Z}\,
\end{equation}
applying it to \eqref{cont} and retaining all distributional terms. Clearly, a 
mathematically sound treatment of the transformation \eqref{trans} is a delicate 
matter. 

A first rigorous result in this realm has been established in
\cite{KS:99a} in the special case of impulsive \emph{pp}-waves, i.e.,
nonexpanding impulsive waves propagating in a Minkowski background, hence 
$\Lambda=0$ in the above metrics \eqref{cont}, \eqref{confppimp}. In 
particular, nonlinear distributional geometry (\cite[Chapter 3]{GKOS:01}) based on 
algebras of generalized functions (\cite{C:85}) has been employed to show the 
following: The `discontinuous coordinate change' \eqref{trans}
(cf.\ \cite{Pen:72} for the plane wave and \cite{AB:97,PV:98} for the general 
\emph{pp}-wave case) relating the distributional Brinkmann form of the metric, 
i.e., \eqref{confppimp} with $\Lambda=0$ 
to the continuous Rosen form, i.e., \eqref{cont} with $\Lambda=0$ 
is the distributional limit of a `generalized diffeomorphism'. For details on 
the latter concept see \cite{EG:11,EG:13}. This result rests on the
nonlinear distributional analysis of the geodesics in the metric 
\eqref{confppimp} with $\Lambda=0$, providing an existence, uniqueness and 
completeness result (\cite{S:98,KS:99}). 

Observe that especially the completeness result is remarkable since it proves 
that the analyti\-cally `very singular' distributional spacetime is 
nonsingular in view of the standard definition (\cite{Pen:65}). More precisely,  
the metric \eqref{confppimp} possesses a distributional coefficient hence lies 
outside the Geroch-Traschen class of metrics (\cite{GT:87}). Indeed this class 
is defined by the metric being of Sobolev regularity $W^{1,2}_{\mbox{\small 
loc}}\cap L^\infty_{\mbox{\small loc}}$ and is known to be the most general 
class which allows to stably define the curvature in distributions (see also 
\cite{LFM:07,SV:09}). Concerning the use of geodesic completeness as a 
criterion for a spacetime to be nonsingular observe that the metric 
\eqref{confppimp} is of regularity far below the class, which classically 
guarantees even the \emph{local existence} of geodesics, which is $C^1$. Still 
more, it is outside the class of metrics which classically allows for 
\emph{unique} local solvability of the geodesic equation ($C^{1,1}$, i.e., the 
first derivatives of the metric being locally Lipschitz continuous), which is the 
natural regularity assumption for the singularity theorems whose validity has 
recently been extended to this class (\cite{KSSV:15,KSV:15}). From this point of 
view it becomes quite clear that in this low regularity scenario both the 
solution concept for the geodesic equation as well as the notion of geodesic 
completeness has to be severely generalized as compared to the classical 
context and we will explain the respective notions in Section~\ref{sec:col}, 
below.
\medskip

A reasonable first step in the long term project to rigorously treat the 
`discontinuous transformation' \eqref{trans} also in the case of nonvanishing 
$\Lambda$ is a nonlinear distributional analysis of the geodesics of 
the metric \eqref{confppimp}. For such an analysis it turns out that a
five-dimensional approach, i.e., describing the (anti-)de Sitter space with 
impulsive wave as a hyperboloid in a five-dimensional flat space with 
impulsive wave (\cite{PG:98,PG:99}), is much better suited than a direct 
approach, as can be seen from the wild singularities encountered in the geodesic 
equations in \cite[Appendix C]{Sfet}. 
Indeed in \cite{PO:01} the five-dimensional 
formalism  has been used to \emph{formally} derive the geodesics of 
all nonexpanding impulsive waves propagating in constant curvature backgrounds 
with $\Lambda\not=0$ and to explicitly relate these geodesics to the geodesics 
of the respective background. More precisely in the five-dimensional approach 
one describes the impulsive spacetime as the (anti-)de Sitter hyperboloid given 
by
\begin{equation}\label{const}
Z_{2}^{2}+Z_{3}^{2}+ \sigma Z_{4}^{2}-2UV=\sigma a^{2},
\end{equation}
with $a:=\sqrt{3/(\sigma \Lambda)},\ \sigma:=\text{sign}(\Lambda)=\pm 1$
in the five-dimensional impulsive \textit{pp}-wave manifold
\begin{equation}\label{classical}
\d s^{2}=\d Z_{2}^{2}+\d Z_{3}^{2}+ \sigma
\d Z_{4}^{2}-2\d U \d V +H(Z_{2},Z_{3},Z_{4})\delta(U)\d U^{2}.
\end{equation}
Here $(Z_0,\dots,Z_4)$ are global Cartesian coordinates of $\R^5$ and ${U=\frac{1}{\sqrt2}(Z_0+Z_1)}$, 
$V=\frac{1}{\sqrt2}(Z_0-Z_1)$ are null-coordinates, which are different from those used in 
the metric \eqref{cont}. Since in this paper we will not use the continuous form \eqref{cont} we simplify the notation by not 
distinguishing them by a bar (as done in \cite{PSSS:15}).

In part I of this series of papers \cite{SSLP:16} we have combined this 
five-dimensional formalism together with a careful regularization approach to 
provide a rigorous analysis of the geodesics. More precisely, we have 
replaced the five-dimensional impulsive \emph{pp}-wave \eqref{classical} by 
the regularized spacetime $(\bar M=\R^5,\bar g_\eps)$ with line element    
\begin{equation}\label{5ipp}
\d\bar s_{\eps}^{2}=\d Z_{2}^{2}+\d Z_{3}^{2}+ \sigma \d Z_{4}^{2}-2\d U \d V
 +H(Z_{2},Z_{3},Z_{4})\delta_{\eps}(U) \d U^{2}\,.
\end{equation}  
Here $\eps\in(0,1]$ is fixed and $\delta_\eps$ is a \emph{model delta net} 
constructed from an arbitrary smooth function $\rho$ on $\R$ with unit integral 
and support in $[-1,1]$ by setting $\delta_\eps(x):=(1/\eps)\,\rho(x/\eps)$. 
Hence $(\bar M,\bar g_\eps)$ may be viewed as a \emph{sandwich wave} which is 
flat space outside the \emph{wave zone} given by $|U|\leq\eps$. The regularized 
impulsive wave spacetime of our interest $(M,g_\eps)$ is now given by the 
(anti-)de Sitter hyperboloid \eqref{const} embedded in $(\bar M,\bar g_\eps)$. 
The main achievement of part I on the geodesics in $(M,g_\eps)$ can now be 
formulated: We consider any geodesic $\gamma=(U,V,Z_p)$ in the regularized 
nonexpanding impulsive wave spacetime $(M,g_\eps)$ starting \emph{outside} the 
wave zone $|U|\leq \eps$ and heading towards it. Before $\gamma$ hits the wave 
zone (for the first time), i.e., $U=-\eps$  it is also a geodesic of the 
background, called \emph{seed geodesic}, and we can fix some 
($\eps$-independent) initial data for $\gamma$ such that its speed is 
normalized, i.e., $|\dot\gamma|=e=\pm 1,0$. When the seed geodesic hits the wave 
zone it continues at least for some small parameter time as a local solution 
$\gamma_\eps$ of the regularized geodesic equations. Now \cite[Theorems 3.1 and 
3.2]{SSLP:16} guarantee that there exists $\eps_0$ such that $\gamma_\eps$ 
passes through the wave zone to the background region `behind' it, 
\emph{provided   $\eps\leq \eps_0$}. 

Here \label{page:discussion} $\eps_0$ depends on 
the seed $\gamma$ via its initial data, cf.\ the discussion in part I, p.\ 11. 
Unfortunately this fact obstructs a straight forward formulation of 
completeness results: Although we can give precise bounds on $\eps_0$ in terms 
of the seed geodesics' data (see \cite[Equation (A.14)]{SSLP:16}) there is no 
global $\eps$ such that all geodesics are complete. Consequently none of the 
spacetimes $(M,g_\eps)$ is geodesically complete (in the usual sense of the 
definition). Even worse, geodesics in the background spacetime with $\sigma
e>0$ (spacelike geodesics in de Sitter and timelike geodesics in
anti-de Sitter space) are periodic. Consequently geodesics $\gamma_\eps$ in the
regularized spacetime constructed from such seed geodesics $\gamma$ will
cross the wave zone infinitely often and we have to reapply the above results 
repeatedly with changing initial data. While this allows us to specify an 
$\eps_0$ which will guarantee that $\gamma_\eps$ crosses the wave zone $N$ 
times for any fixed integer $N$, we cannot give a global $\eps_0$ for which the 
geodesic $\gamma_\eps$ is complete. 
\medskip

In this paper we remedy these defects by invoking the geometric theory of 
generalized functions (\cite[Chapter 3]{GKOS:01}) which will allow us to formulate 
a \emph{`global' completeness statement}. We will show global existence and 
uniqueness of geodesics in this framework and, in particular, handle the case of 
infinitely many crossings thereby establishing completeness for all geodesics in 
nonexpanding impulsive gravitational waves with $\Lambda$, which is also a 
completely new result as compared to \cite{SSLP:16}.

Geodesic completeness of (semi-)Riemannian manifolds in the nonlinear 
distributional setting has been first considered explicitly in \cite{SS:15}, 
where also completeness of a different class of impulsive gravitational waves 
has been established. This class consists of impulsive versions of models 
considered in \cite{CFS:03,FS:03,CFS:04,FS:06}  and called (general) 
plane-fronted waves (PFW) there. They generalize \emph{pp}-waves by allowing for 
an arbitrary $n$-dimensional Riemannian manifold $N$ as wave surface and 
consequently may be called \emph{$N$-fronted waves with parallel rays (NPWs)}. 
Now the geodesics in impulsive NPWs (iNPWs) have been analyzed again using a 
regularization approach in \cite{SS:12}, which was turned into a 'global' 
completeness result (in the sense discussed above) in \cite{SS:15}. Our current 
approach in some respect parallels the one of \cite{SS:12,SS:15}, however the 
technical complexity is severely increased, cf.\ \cite[p.\ 19]{SSLP:16}.
\medskip

This paper is organized as follows. To keep our presentation self contained we 
briefly recall the basic elements of nonlinear distributional geometry in 
Section~\ref{sec:col} where we also transfer the five-dimensional formalism to 
the generalized setting. We prove global existence and uniqueness to the 
system of geodesic equations in generalized functions in Section 
~\ref{sec:complete}. Our main result on completeness of nonexpanding 
impulsive gravitational waves in (anti-)de Sitter universe is given in 
Section \ref{sec:compl1}. Finally we explicitly relate these complete 
geodesics to the geodesics of the cosmological background in section 
\ref{sec:4}.



\section{Nonlinear distributional geometry}\label{sec:col}

To keep this presentation self-contained, we begin this section with a brief  
review of generalized semi-Rie\-man\-nian geometry in the setting of 
Colombeau's nonlinear generalized functions. Colombeau algebras of generalized 
functions (\cite{C:85}) 
are differential algebras which contain the vector space of 
Schwartz distributions and at the same time display maximal consistency with 
classical analysis. The theory of semi-Riemannian geometry based on the 
so-called special Colombeau algebra $\G(M)$ has been 
developed in \cite{KS:02a,KS:02b}, see also \cite[Section 3.2]{GKOS:01}. 
The basic idea of the construction to be detailed below 
is regularization of distributions via nets of smooth functions combined with 
an elaborate bookkeeping of asymptotic estimates in terms of a regularization 
parameter. 

Let $M$ be a smooth (second countable and Hausdorff) manifold  and denote by 
$\E(M)$ the set of all nets $(u_\eps)_{\eps\in (0,1]=:I}$ in $\Cinf(M)^I$ 
depending smoothly on $\eps$. Smooth dependence on the parameter renders 
the theory technically more pleasant but was not assumed in earlier references, 
cf.\ the discussion in \cite[Section 1]{BK:12}. The \emph{algebra of generalized 
functions on $M$} (\cite{DD:91}) is defined as the quotient $\G(M) := 
\EM(M)/\NN(M)$ of \emph{moderate} modulo \emph{negligible} nets in $\E(M)$, 
which are defined via 
\[
\EM(M) :=\{ (u_\eps)_\eps\in\E(M):\, \forall K\Subset M\
\forall P\in{\mathcal P}\ \exists N:\ \sup\limits_{p\in 
  K}|Pu_\eps(p)|=O(\eps^{-N}) \},\\
\]
\[\NN(M)  :=\{ (u_\eps)_\eps\in\EM(M):\ \forall K\Subset M\
\forall m:\ \sup\limits_{p\in K}|u_\eps(p)|=O(\eps^{m}) \},
\]
where ${\mathcal P}$ denotes the space of all linear differential operators on 
$M$ and $K\Subset M$ means that $K$ is a compact subset of $M$. Elements of $\G(M)$ are denoted by $u = 
[(u_\eps)_\eps]$ and we call $(u_\eps)_\eps$ a representative of
the generalized function $u$.
Defining sum and product in $\G(M)$ componentwise (i.e., for $\eps$ fixed) and 
the Lie derivative with respect to smooth vector fields $\xi\in{\mathfrak X(M)}$ via 
$L_\xi u :=[(L_\xi u_\eps)_\eps]$, $\G(M)$ becomes a \emph{fine sheaf of 
differential algebras}.

There exist embeddings $\iota$ of the space of distributions $\D'(M)$ into 
$\G(M)$ that are sheaf homomorphisms and preserve the product of 
$\Cinf(M)$-functions. A coarser way of relating generalized functions in 
$\G(M)$ to distributions is as follows: $u\in \G(M)$ is called 
\emph{associated} with $v\in \G(M)$, $u\approx v$, if $u_\eps - v_\eps \to 0$
in $\D'(M)$. Moreover, $w\in \D'(M)$ is called associated with $u$ if 
$u\approx \iota(w)$.

The ring of constants in $\G(M)$ is the space $\tilde \R = \EM / \NN$ of generalized
numbers, which form the natural space of point values
of Colombeau generalized functions. These, in turn, are uniquely
characterized by their values on so-called compactly supported
generalized points.
\smallskip

More generally the space $\GaG(M,E)$ of \emph{generalized sections}  
of the vector bundle $E$ is defined as
$
  \label{tensorp} \GaG(M,E) = \G(M) \otimes_{\Cinf(M)} \Ga(M,E)= 
  L_{\Cinf(M)}(\Ga(M,E^*),\G(M)).
$ 
It is a fine sheaf of finitely generated and projective $\G$-modules. 
For \emph{generalized tensor fields} of rank $r,s$ we use 
the notation $\G^r_s(M)$, i.e.\ 
\begin{equation}
  \G^r_s(M)\cong L_{\G(M)}(\G^1_0(M)^s,\G^0_1(M)^r;\G(M)). 
\end{equation}
Observe that this, in particular, allows the insertion of generalized vector 
fields and one-forms into generalized tensors, which is not possible in the 
distributional setting (cf.\ \cite{marsden,deR}) but essential when dealing 
with generalized metrics. Here a \emph{generalized pseudo-Riemannian metric} is 
a section $g\in\G^0_2(M)$ that is symmetric with determinant $\det g$ 
invertible in $\G$ (equivalently $|\det (g_\eps)_{ij}| > \eps^m$ 
for some $m$ on compact sets), and a well-defined 
index $\nu$ (the index of $g_\eps$ equals $\nu$ for $\eps$ small). By a 
``globalization Lemma'' in (\cite[Lemma 2.4, p.\ 6]{KSSV:14}) any generalized metric 
$g$ possesses a representative $(g_\eps)_\eps$ such that each $g_\eps$ is a 
smooth metric globally on $M$. We call a pair $(M,g)$ consisting of a smooth 
manifold and a generalized metric a generalized semi-Riemannian manifold.

Based on this definition a convenient framework for non-smooth pseudo-Riemannian 
geometry has been developed. It in turn enables an analysis of spacetimes of low 
regularity in general relativity consistently extending the ``maximal 
distributional'' setting of Geroch and Traschen (\cite{GT:87}), see 
\cite{SV:09,S:08}. In particular, any generalized metric induces an isomorphism 
between generalized vector fields and one-forms, and there is a unique 
Levi-Civita connection $\nabla$ corresponding to $g$.
\smallskip

Finally to discuss geodesics in generalized semi-Riemannian manifolds
we have to introduce the space of generalized curves defined on an 
interval $J$ \emph{taking values in the manifold $M$} $\G[J,M]$. It is again a 
quotient of moderate modulo negligible nets $(\gamma_\ep)_\ep$ of smooth curves, 
where we call a net moderate (negligible) if $(\psi\circ \gamma_\eps)_\eps$ is 
moderate (negligible) for all smooth $\psi:M\to\R$. In addition, 
$(\gamma_\eps)_\eps$ is
supposed to be \emph{c-bounded}, which means that $\gamma_\eps(K)$ is contained in a compact
set for $\eps$ small and all compact sets $K\Subset J$. Observe that no distributional
counterpart of such a space exists.

The \emph{induced covariant derivative} of a generalized vector field $\xi$ on 
a generalized curve $\gamma=[(\gamma_\eps)_\eps]\in\G[J,M]$ 
is defined componentwise and gives again a generalized vector field $\xi'$ 
on $\gamma$. In particular, a \emph{geodesic} in a generalized 
pseudo-Riemannian manifold is a curve $\gamma\in\G[J,M]$ satisfying 
$\gamma''=0$. Equivalently the usual local formula holds, i.e.,  
\begin{equation}\label{geo}
  \Big[\,\Big(\frac{d^2\gamma_\eps^k}{dt^2}
  +\sum_{i,j}\Ga^k_{\eps 
ij}\frac{\gamma_\eps^i}{dt}\frac{\gamma_\eps^j}{dt}\Big)_\eps\,\Big]
  =0,
\end{equation}
where $\Ga^k_{ij}=[(\Ga^k_{\eps ij})_\eps]$ denotes the Christoffel symbols of 
the generalized metric
$g=[(g_\eps)_\eps]$. Now the following definition is natural.
\begin{defi}\label{def-geo-compl} (\cite[Definition 2.1, p.\ 240]{SS:15})
Let $g\in\G_0^2(M)$ be a generalized semi-Riemannian metric. Then $(M,g)$ is said to be 
\emph{geodesically complete} if every geodesic $\gamma$ can be defined on $\R$, i.e., every solution of the geodesic equation
\begin{equation}
 \gamma''=0,
\end{equation}
is in $\G[\R,M]$.
\end{defi}
\smallskip

Now we set up the five-dimensional framework to deal with impulsive waves 
propagating in cosmological backgrounds. To begin with we recall that a
\emph{model delta function} is an element $D\in\G(\R)$ that has as a 
representative a \emph{model delta net} $\delta_\eps(x):= 
\frac{1}{\eps}\rho(\frac{x}{\eps})$, where again
$\rho\in\Cinf(\R)$ with compact support in $[-1,1]$ and $\int_\R \rho = 1$.

Now let $H\in\Cinf(\R^3)$ and $\sigma := \text{sign}(\Lambda) = \pm 1$. Then we 
define the $5$-dimensional generalized impulsive \emph{pp}-wave manifold 
$(\bar M=\R^5,\bar g)$  via 
\begin{equation}\label{eq:Mbar}
  \d\bar s^2= dZ^2_2 + dZ^2_3 + \sigma dZ^2_4 - 2 dU dV + H(Z_2,Z_3,Z_4)D(U) 
dU^2\,,
\end{equation}
where $Z_2, Z_3, Z_4$ are global Cartesian coordinates on $\R^3$ and $U, V$ are 
global null coordinates. One easily checks that this defines a generalized 
metric in the sense above. At this point we specify the hypersurface $M$ in 
$(\bar M,\bar g)$, which will be 
our main playground
\begin{align}\label{eq:M}\nonumber
  M := & \{(U,V,Z_2,Z_3,Z_4)\in\bar M : F(U,V,Z_2,Z_3,Z_4)=0\}\,,\quad 
\mbox{where}\\
&F(U,V,Z_2,Z_3,Z_4):=-2UV+Z_{2}^{2}+Z_{3}^{2}+ 
\sigma Z_{4}^{2}-\sigma a^{2}\,.
\end{align}
Note that $M$ is a (classical) smooth hypersurface. Finally we restrict 
the metric $\bar g$ to $M$ to obtain the generalized spacetime $(M,g)$ which 
we take as our model of nonexpanding impulsive waves propagating on a(n 
anti-)de Sitter universe.
%
%

To derive the geodesic equations in $(M,g$) we recall that all the classical formulas hold
componentwise (i.e., for fixed $\eps$) in nonlinear generalized functions. So we may use the classical 
condition that the geodesics' $\bar M$-acceleration is normal to $M$,
$\bar\nabla_T T=-\sigma g(T,\bar\nabla_T N)N/g(N,N)$ to derive the geodesic equations. Here $T$ 
is the geodesic tangent and $N$ 
denotes the (non-normalized) normal vector to $M$ defined via its representative 
$N^\alpha_\eps=g^{\alpha\beta}_\eps dF_\beta$, cf.\ \cite[p.\ 6]{SSLP:16}.  In 
this way we 
arrive for the representatives $\gamma_\eps=(U_\eps,V_\eps,Z_{p\eps})$ of a generalized
geodesic $\gamma=(U,V,Z_p)$ in $M$ precisely at \cite[(2.17)]{SSLP:16}, which 
for $\gamma$ gives
\begin{align}\label{eq:geos:G}
  \ddot{U}
  &=-\Big( e + \frac{1}{2}\,\dot{U}^2\,\tilde{G}
  - \dot{U}\,\big(H\,D\,U\dot{\big)}\Big)\
  \frac{U}{\sigma a^2-U^2 H D}\,, \nonumber\\
  \ddot{V}-\frac{1}{2}\,H\,\dot{D}\,\dot{U}^2 - 
  \delta^{pq}H_{,p}\,\dot{Z}_q\,D\,\dot{U}
  &=-\Big(e + \frac{1}{2}\,\dot{U}^2\,\tilde{G}- \dot{U}\,\big( H\, D\, U 
  \dot{\big)}\Big)\
  \frac{V+H\,D\, U}{\sigma a^2-U^2 H D}\,,\nonumber\\
  \ddot{Z}_{i}-\frac{1}{2}H_{,i}\,D\, \dot{U}^2 &=-\Big(e + 
  \frac{1}{2}\,\dot{U}^2\,\tilde{G}
  - \dot{U}\,\big( H\, D\, U \dot{\big)}\Big)\ \frac{Z_{i}}{\sigma 
    a^2-U^2 H D}\,,\\
  \ddot{Z}_{4}-\frac{\sigma}{2}\,H_{,4}\,D\,\dot{U}^2 &=-\Big(e + 
\frac{1}{2}\,\dot{U}^2\,\tilde{G}
          - \dot{U}\,\big( H\, D\, U \dot{\big)}\Big)\
    \frac{Z_{4}}{\sigma a^2-U^2 H D}\,. \nonumber
\end{align}

Here again $e=|\dot\gamma|=\pm 1,0$ for which it is natural to be fixed 
independently of $\eps$.

\section{Unique global existence of geodesics crossing the 
impulse}\label{sec:complete}

In this section we prove existence and uniqueness of globally defined 
(generalized) solutions of the geodesic equations \eqref{eq:geos:G} with 
suitable initial data that enforces them to cross the wave impulse.

To begin with we recall the basic steps one has to take to solve an initial 
value problem (IVP) in generalized functions. We consider
\begin{equation}\label{ivp-G}
\begin{cases}
 &\dot{Z}(t) = F(Z(t))\,,\\
 &Z(t_0) = Z^0\,,
\end{cases}
\end{equation}
where we assume the right hand side to be a generalized function 
$F\in\G(\R^n,\R^n)$ and the initial condition to be a generalized 
vector $Z^0\in\Rt^n$. We look for solutions $Z\in\G[J,\R^n]$ on an interval 
$J\subseteq\R$ containing $t_0$. For all details we refer to \cite{KOSV:04}.

To prove existence and uniqueness of \eqref{ivp-G} one basically solves the 
IVP componentwise, i.e., for each $\eps$ one solves 
\begin{equation}\label{ivp-G-eps}
\begin{cases}
 &\dot{Z_\eps}(t) = F_\eps(Z_\eps(t))\,,\\
 &Z_\eps(t_0) = Z_\eps^0\,,
\end{cases}
\end{equation}
where $F=[(F_\eps)_\eps]$, $Z=[(Z_\eps)_\eps]$ and $Z^0 = [(Z_\eps^0)_\eps]$ and then derives the
necessary asymptotic estimates. In some more detail one proceeds in the following three steps.
\begin{enumerate}
 \item First, one proves existence of a so-called \emph{solution candidate}, i.e., 
a net of smooth functions $Z_\eps\colon J\rightarrow \R^n$ depending smoothly on 
the parameter $\eps$ such that for 
all fixed (and small) $\eps$ we have $\dot Z_\eps = F_\eps(Z_\eps)$, 
$Z_\eps(t_0)=Z^0_\eps$. Note that the domain of definition
$J$ of the solution candidate has to be independent of $\eps$ or at least 
to depend on $\eps$ favorably (e.g.\ not shrinking to a point for 
$\eps\to 0$).
\item Second, one shows \emph{existence} of a generalized solution by establishing c-boundedness and moderateness of the 
solution candidate, i.e., $Z:=[(Z_\eps)_\eps]\in\G[J,\R^n]$.
\item Third, to show \emph{uniqueness} in $\G$ one solves a negligibly perturbed version 
of \eqref{ivp-G-eps}, i.e., $\dot{\tilde Z}_\eps = F_\eps(\tilde 
Z_\eps)+a_\eps$, $\tilde Z_\eps(t_0)=Z^0_\eps+b_\eps$ with negligible 
$(a_\eps)_\eps$, $(b_\eps)_\eps$ and shows that the corresponding net of 
solution $(\tilde{Z}_\eps)_\eps$ only differs negligibly from $Z$, i.e., 
$[(\tilde{Z}_\eps)_\eps] =[(Z_\eps)_\eps] = Z$.
\end{enumerate}
Note that a uniqueness result in $\G$ amounts to an additional stability 
result, since it says that negligible perturbations of the 
initial value problem lead to only negligibly perturbed solutions. 
\medskip

In our present analysis we aim at establishing geodesic completeness of the 
impulsive wave spacetime $(M,g)$ with generalized metric $g$. So we have to 
prove that all solutions of the geodesic equations \eqref{eq:geos:G} are global. 
In this section we will prove global existence and uniqueness of solutions 
given any initial data that forces them to run into the impulsive wave. For 
convenience we will generate such arbitrary initial data from geodesics of the 
constant curvature background, cf.\ the discussion of seed geodesics in Section 
\ref{sec:intro}. All other geodesics, i.e., those which do not cross the 
impulsive wave are simple to deal with and we will briefly do so in the 
following section prior to presenting the completeness result.

The main issue in showing \emph{global} existence is that we have to
first construct a global solution candidate. To this end we will use the 
power of the theory of nonlinear generalized functions to overcome the obstacle 
that the regularization approach in combination with the fixed point argument 
detailed in part I does not produce a net of global $\eps$-wise 
solutions, cf.\ the discussion on page \pageref{page:discussion}. Indeed the 
$\eps_0$ from which on a geodesics is complete depends in its initial data. 
This, in particular, affects the case $\sigma e>0$ where the geodesics cross 
the wave impulse arbitrarily often. Observe that in our approach it is 
a \emph{result} that the geodesics preserve the norm of their tangent vector 
when crossing the impulse (cf.\ \cite[Remark 3.3]{SSLP:16}) and hence their 
causal character. As discussed in the introduction (see also \cite[p.\ 
10f.]{SSLP:16}) the `pure' regularization approach of part I does not allow to 
prove existence for infinitely many crossings.  Here, however, we handle also 
this case when constructing a global solution candidate.
  
We will nevertheless start by first constructing a locally defined solution 
candidate, i.e., a net of geodesics that cross the wave impulse 
\emph{once} and we will only later extend it to a global solution candidate.

\subsection{Construction of a local solution candidate}\label{subsec:sol-cand}

Already in the construction of a local solution candidate, to be detailed in 
this subsection, we need to generalize the approach of \cite{SSLP:16} 
such as to allow the initial data to be generated by a \emph{family} of seed 
geodesics  rather than by a single seed geodesic.

To begin with consider a family of geodesics 
$\gamma^-_\eps=(U^-_\eps,V^-_\eps,Z^-_{p\eps})$ of the background (an\-ti-)\-de 
Sitter universe \emph{without} impulsive wave but reaching $U=0$ (all other 
geodesics are not of interest now and will be dealt with separately in section 
\ref{sec:compl1}). Without loss of generality we can assume that $U^-_\eps(0)=0$ 
by choosing an affine parameter for $\gamma^-_\eps$ appropriately and we assume 
$\dot \gamma^-_\eps$ to be normalized by $e=\pm 1,0$, independently of $\eps$. 
Furthermore, again without loss of generality we can assume that the $U$-speed 
when reaching $U=0$, call it $\dot U^{-0}_\eps$, is positive (the case $\dot 
U^{-0}_\eps<0$ can be treated in complete analogy) so that locally $U^-_\eps$ is 
increasing. It is thus most convenient to prescribe initial data at the affine 
parameter value $0$, that is we set 
\begin{equation}\label{eq:data0}
 \gamma^-_\eps(0)=:(0,V^0_\eps,Z^0_{p\eps})\,,\qquad \dot \gamma^-_\eps(0)=:(\dot U^{0}_\eps>0,\dot V^0_\eps,\dot 
Z^0_{p\eps})\,,
\end{equation}
where the constants satisfy the constraints
\begin{equation}\label{const1}
(Z^0_{2\eps})^{2}+(Z^0_{3\eps})^{2}+ \sigma (Z^0_{4\eps})^{2}=\sigma a^{2},\quad
Z^0_{2\eps} \dot Z^0_{2\eps}+Z^0_{3\eps} \dot Z^0_{3\eps}+ \sigma Z_{4\eps}^0\dot  Z_{4\eps}^0-V^0_\eps\dot U^0_\eps=0\,,
\end{equation}
for every $\eps$, which follows from \eqref{const} (with $U^0_\eps=0$). In addition the normalization
condition
\begin{equation}\label{eq:norm}
 -2 \dot U^0_\eps\dot V^0_\eps+(\dot Z^0_{2\eps})^2+(\dot Z^0_{3\eps})^2+\sigma(\dot Z^0_{4\eps})^2=e
\end{equation}
holds for every $\eps$. Now we assume that the net $(\gamma^-_\eps(0))_\eps$ of 
data at $0$ converges, more precisely we assume
\begin{equation}\label{eq:data0-conv}
 \lim_{\eps\searrow 0} \gamma^-_\eps(0) =: (0,V^0,Z^0_p)\,,\qquad \lim_{\eps\searrow 0}\dot 
\gamma^-_\eps(0)=:(\dot U^0>0,\dot V^0,\dot Z^0_p)\,,
\end{equation}
which automatically satisfy the constraint and the normalization. We will refer 
to a family $(\gamma^-_\eps)_\eps$ as described here as a \emph{family of seed 
geodesics}. Note that $(\gamma^-_\eps)_\eps$ depends on $\eps$ only via its data 
and that it converges locally uniformly together with its first derivative to a 
fixed background geodesic $\gamma^-$ with data \eqref{eq:data0-conv} by 
continuous dependence of solutions of ODEs on their data.
\medskip

At this point we start to think of $\gamma^-_\eps$ as geodesics in the 
impulsive wave spacetime \eqref{classical}, \eqref{const} \emph{`in front' of 
the impulse} that is for $U^-_\eps<0$. Also, $\gamma^-_\eps$ are geodesics in 
the regularized spacetime \eqref{5ipp},
\eqref{const} \emph{`in front' of the sandwich wave}, that is for $U^-_\eps\leq
-\eps$. We will denote the affine parameter time when $\gamma^-_\eps$ enters 
this regularization wave region by $\alpha_\eps$,
\begin{equation}
 U^-_\eps(\alpha_\eps)=-\eps\,.
\end{equation}
Observe that $\alpha_\eps\to 0$ from below as $\eps\to 0$. 

Finally we come to set up the data for the solution candidate $\gamma_\eps$ of 
the system \eqref{eq:geos:G} by
\begin{equation}\label{eq:real-data}
 \gamma_\eps(\alpha_\eps)=\gamma^-_\eps(\alpha_\eps),\quad
 \dot \gamma_\eps(\alpha_\eps)=\dot\gamma^-_\eps(\alpha_\eps),
\end{equation}
i.e., as the data the family of seed geodesics assumes at $\alpha_\eps$. We 
will frequently refer to these data \eqref{eq:real-data} as initial data 
constructed from the seed family with data \eqref{eq:data0}.
\bigskip

Now we will provide the local solution candidate by proving that we can extend 
the background geodesics not only into but even \emph{through} the entire wave 
zone. Local existence of such an extension for one crossing is provided by a 
fixed point argument. More precisely it suffices to study the model system 
below. Note that compared to \cite[Equation (A.1), p.\ 20]{SSLP:16} we deal 
with a negligibly perturbed system which is necessary to prove uniqueness in 
Subsection \ref{subsec:uniq}, below. We consider
\begin{align}\label{eq-ivp-eps}
 \ddot{u}_\eps &= -\frac{e u_\eps + \Delta_\eps u_\eps}{N_\eps} + a_\eps\,, \nonumber\\
 \ddot{z}_\eps-\frac{1}{2}DH\,\delta_{\eps}\dot{u}_\eps^2
&=-\frac{e z_\eps + \Delta_\eps z_\eps}{N_\eps} + c_\eps\,,\\
u_\eps(\alpha_\eps)=-\eps,\quad \dot{u}_\eps(\alpha_\eps) &= \dot{u}^0_\eps +d_\eps,\quad z_\eps(\alpha_\eps) = z^0_\eps + 
f_\eps,\quad \dot{z}_\eps(\alpha_\eps) = \dot{z}^0_\eps+h_\eps\,,\nonumber
\end{align}
where
\begin{align*}
 \Delta_\eps &= \frac{1}{2}\,\dot{u}_\eps^2\,{G_\eps}- \dot{u}_\eps\,\big(H 
\,\delta_{\eps}
           \,u_\eps \big)\dot\,,\\
 N_\eps &= \sigma a^2-u_\eps^2H\delta_\eps\,,
\end{align*}
and $H=H(z_\eps)$ is a smooth function on $\R^3$, $DH$ denotes its gradient, and
${G}_\eps(u_\eps,z_\eps):=DH(z_\eps)\,\delta_{\eps}(u_\eps)\,z_\eps +
H(z_\eps)\,\delta'_\eps(u_\eps)\,u_\eps$. Moreover we assume 
$(a_\eps)_\eps\in\NN(\R)$, $(c_\eps)_\eps\in\NN(\R^3)$, $(d_\eps)_\eps\in\NN$, 
$(f_\eps)_\eps, (h_\eps)_\eps\in\NN^3$.
Finally, the initial conditions are specified in the spirit of the seed 
family approach as follows: We fix $\dot{u}^0>0$ and $z^0, \dot{z}^0\in\R^3$ 
and assume 
\begin{equation}\label{eq:model-data}
 x^0_\eps :=(-\eps,z^0_\eps)\to (0,z^0)=:x^0,\quad
 \dot{x}^0_\eps :=(\dot{u}^0_\eps,\dot{z}^0_\eps)
  \to(\dot u^0,\dot z^0)=:\dot x^0
\end{equation}
as $\eps\searrow 0$. Furthermore let 
$\alpha_\eps < 0$ such that $\alpha_\eps\nearrow 0$ for $\eps \searrow 0$. Also 
we will frequently use the notation $x_\eps:=(u_\eps,z_\eps)$.
\medskip

At this point we set up the space of possible solutions. We will outline 
the basic steps of the construction extending the approach of \cite[Appendix 
A]{SSLP:16}.
Let $C_1>0$ and set 
\begin{align*}
 C_2 &:= 1 + \max \Bigl( \frac{81}{2}\|DH\| \|\rho\| \dot{u}^0,\ \frac{12}{a^2}(|z^0|+C_1),\\
 &\frac{54}{a^2}(|z^0|+C_1)\bigl(\frac{3}{2}\dot{u}^0\|DH\|\|\rho\|(|z^0|+C_1) + \frac{3}{2}\dot{u}^0\|H\|\|\rho'\| + 
2\|DH\|\|\rho\| + 3\dot{u}^0\|H\|\|\rho\|\bigr)\Bigr)\,,
\end{align*}
where $\|H\|=\|H\|_{L^\infty(B_{C_1}(z^0))}$ and $\|DH\|=\|DH\|_{L^\infty(B_{C_1}(z^0))}$, the $L^\infty$-norm
on the closed ball of radius $C_1$ around $z^0$.

Now we define $\eta>0$ --- the length of the solution interval (which is independent of $\eps$) by
\begin{align*}
 \eta&:=\min\Bigl\{1,\, \frac{a^2}{4(1+9\dot{u}^0)}, \frac{6 C_1}{25 + 9\dot{u}^0},\,\frac{C_1}{16 
|\dot{z}^0|},\,\frac{C_1}{54 \|DH\|\|\rho\|\dot{u}^0},\,\frac{C_1 a^2}{16(|z^0|+C_1)},\, \frac{C_1 a^2}{24(|z^0|+C_1)}\\
&\times\Bigl(\frac{9}{2}\dot{u}^0\|DH\|\|\rho\|(|z^0|+C_1) + \frac{9}{2}\dot{u}^0\|H\|\|\rho'\|+ 
6\|DH\|(|\dot{z}^0|+C_2)\|\rho\|+\frac{9}{2}\dot{u}^0\|H\|\|\rho\|\Bigr)^{-1}\Bigr\}\,.
\end{align*}

Set $J_\eps:=[\alpha_\eps,\alpha_\eps + \eta]$ and define a closed subset of the complete metric space 
$\Con^1(J_\eps,\R^4)$ by setting
\begin{align}\label{def-Xe}
 \Xe :=  \Bigl\{ x_\eps:=(u_\eps,z_\eps)\in\Con^1(J_\eps, \R^4):\ & 
x_\eps(\alpha_\eps)=x^0_\eps + (0,f_\eps),\ 
\dot{x}_\eps(\alpha_\eps) = \dot{x}^0 + (d_\eps,h_\eps),\nonumber\\ 
&\|x_\eps - x^0\|\leq C_1,\ \|\dot z_\eps-\dot{z}^0\|\leq C_2,\ 
\dot{u}_\eps\in[\frac{1}{2}\dot{u}^0,\frac{3}{2}\dot{u}^0] \Bigr\}\,.
\end{align}
Now we define the solution operator $A_\eps = (A_\eps^1,A_\eps^2)$ on $\Xe$:
\begin{align}
 &A_\eps^1(x_\eps)(t):= -\eps + (\dot{u}_\eps^0 + d_\eps)(t-\alpha_\eps) - 
\int_{\alpha_\eps}^t\int_{\alpha_\eps}^s \frac{e u_\eps + \Delta_\eps u_\eps}{N_\eps}\dd r\,\dd s + \int_{\alpha_\eps}^t 
\int_{\alpha_\eps}^s a_\eps\,\dd r\,\dd s\,,\nonumber\\
 &A_\eps^2(x_\eps)(t):= z^0_\eps + f_\eps + (\dot{z}^0_\eps + h_\eps)(t-\alpha_\eps)\nonumber \\
 &\qquad\qquad\quad + \int_{\alpha_\eps}^t 
\int_{\alpha_\eps}^s \frac{1}{2}DH \delta_\eps \dot{u}_\eps^2 - \frac{e z_\eps + \Delta_\eps z_\eps}{N_\eps}\,\dd r\,\dd s 
+\int_{\alpha_\eps}^t\int_{\alpha_\eps}^s c_\eps\,\dd r\,\dd s\,.\label{eq-A_eps}
\end{align}

Next we will show that the operator $A_\eps$ possesses a unique fixed 
point in $\Xe$. 
The first steps are (cf.\ \cite[Lemma A.1 and A.2, p.\ 21]{SSLP:16}): 
\begin{lem}\label{lem-1.-est}\leavevmode
\begin{enumerate}[(i)]
 \item\label{lem-neps-est} Let $z\in B_{C_1}(z^0)$ and $u\in \R$ then if $\eps\leq \frac{a^2}{2\|\rho\|\|H\|}$ we have
\begin{equation}
  \frac{1}{|\sigma a^2 - u^2 H \delta_\eps|}\leq \frac{2}{a^2}\,.
\end{equation}

\item\label{lem-diam} For $x_\eps\in\Xe$ define $\Gamma_\eps(x_\eps):=\{t\in J_\eps: |u_\eps(t)|\leq \eps\}$ to be the 
parameter interval, where $u_\eps$ is in the regularized wave zone, then
 \begin{equation}
  \diam(\Gamma_\eps(x_\eps))\leq\frac{4\eps}{\dot{u}^0}\,.
 \end{equation}
\end{enumerate}
\end{lem}

Now we can quantify how small $\eps$ has to be. To begin with let
\begin{align}\nonumber\label{eeq-eps-0}
 \eps_0':=&\min\Bigl \{\frac{1}{12},\,\frac{1}{12 
a^2}\bigl(\dot{u}^0(\|\rho\|(\frac{3}{2}\|DH\|(|z^0|+C_1)+2\|DH\|(|\dot{z}^0|+C_2)+\frac{3}{2}\|H\|+\frac{3}{2}
\|H\|\|\rho'\|\bigr)^{-1},\\
&\qquad\ \ 
\eta,\,\frac{C_1}{16},\,\frac{C_2}{6},\,\frac{1}{|\dot{z}^0|+C_2},\,\frac{
  \eta \dot u^0}{6}\Bigr\} \ ,
\end{align}
and let $0<\eps_0\leq\eps_0'$ such that for all $0<\eps\leq\eps_0$
\begin{align}\label{eq-eps_0}
 &|d_\eps|\leq\eps,\,\|f_\eps\|\leq\eps,\,\|h_\eps¸\|\leq\eps,\,|a_\eps|\leq\eps 
\text{ (on }[-1,1]\text{)},\,\|c_\eps\|\leq\eps \text{ (on 
}[-1,1]\text{)},\nonumber\\
&|\dot{u}_\eps^0-\dot{u}^0|\leq \frac{1}{8},\,\|z^0_\eps - z^0\|\leq \frac{C_1}{8},\text{ and } 
\|\dot{z}^0_\eps-\dot{z}^0\|\leq \min(\frac{C_1}{16},\frac{C_2}{6})\,.
\end{align}

With these preparations one shows that $A_\eps$ maps $\Xe$ to $\Xe$ 
(cf.\ \cite[Proposition A.3, p.\ 22]{SSLP:16}) as well as the results analogous 
to \cite[Lemma A.4, p.\ 25, Proposition A.5, p.\ 25]{SSLP:16} 
(in fact, with modified constants) providing the necessary estimates which in 
turn allow to show that $A_\eps$ possesses a fixed point in $\Xe$ and hence 
that \eqref{eq-ivp-eps} has a local solution candidate (cf.\ \cite[ Theorem 
A.6, p.\ 27]{SSLP:16}).

\begin{prop}[Local solution candidate]\label{prop:exmat}
Consider the IVP \eqref{eq-ivp-eps} with $\eps\leq\eps_0$ where $\eps_0$ is 
given by \eqref{eeq-eps-0}, \eqref{eq-eps_0}. Then there is a unique smooth 
solution $(u_\eps,z_\eps)$ on $[\alpha_\eps,\alpha_\eps+\eta]$, depending 
smoothly on $\eps$. Moreover, $u_\eps$ and $z_\eps$ as well as their first order 
derivatives are uniformly bounded in $\eps$.
\end{prop}

\begin{pr}
The above construction allows the application of Weissinger's fixed point theorem (\cite{Wei:52}) for fixed $\eps\leq 
\eps_0$ and $\eta$ as above, providing thus a unique fixed point for the operator $A_\eps$ on the space $\mathfrak{X}_\eps$ 
which in turn gives a unique ${\mathcal C}^1$-solution $x_\eps=(u_\eps,z_\eps)$ on $[\alpha_\eps,\alpha_\eps+\eta]$  to the 
IVP \eqref{eq-ivp-eps}. Moreover, since the right hand side of the system 
is smooth the solution is smooth as well, and in addition depends smoothly on 
$\eps$.

The solution obtained via the fixed point argument is unique in the space $\mathfrak{X}_\eps$
and thereby unique among all smooth solutions assuming this data by the usual
argument from ODE-theory.

Finally,  $u_\eps$, $\dot u_\eps$, $z_\eps$, and $\dot z_\eps$ are bounded
uniformly in $\eps$ on $[\alpha_\eps,\alpha_\eps+\eta]$ by the very definition of $\mathfrak{X}_\eps$.
\end{pr}

Going back from the model IVP \eqref{eq-ivp-eps} to the geodesic equations 
\eqref{eq:geos:G} on the level of representatives
\begin{align}\label{eq:geos:eps}
  \ddot{U}_\eps
  &=-\Big( e + \frac{1}{2}\,\dot{U}_\eps^2\,{G_\eps}
  - \dot{U}_\eps\,\big(H
  \,\delta_{\eps}
  \,U_\eps\dot{\big)}\Big)\
  \frac{U_\eps}{\sigma a^2-U_\eps^2H\delta_\eps}\,, \nonumber\\
  \ddot{V}_\eps-\frac{1}{2}\,H
  \,\delta^{'}_{\eps}
  \,\dot{U}_\eps^2 - \delta^{pq}H_{,p}
  \,\delta_{\eps}
  \,\dot{Z}^\eps_q\,\dot{U}_\eps
  &=-\Big(e + \frac{1}{2}\,\dot{U}_\eps^2\,{G}_\eps
  - \dot{U}_\eps\,
  \big( H\, \delta_\eps\, U_\eps \dot{\big)}\Big)\
  \frac{V_\eps+H\,\delta_{\eps}U_\eps}
  {\sigma a^2-U_\eps^2H\delta_\eps}\,,\nonumber\\
  \ddot{Z}_{i\eps}-\frac{1}{2}H_{,i}\,\delta_{\eps}\dot{U}_\eps^2
  &=-\Big(e + \frac{1}{2}\,\dot{U}_\eps^2\,{G}_\eps
  - \dot{U}_\eps\,
  \big( H\, \delta_\eps\, U_\eps \dot{\big)}\Big)\
  \frac{Z_{i\eps}}{\sigma a^2-U_\eps^2H\delta_\eps}\,,\\
  \ddot{Z}_{4\eps}-\frac{\sigma}{2}\,H_{,4}\,\delta_{\eps}\dot{U}_\eps^2
  &=-\Big(e + \frac{1}{2}\,\dot{U}_\eps^2\,{G}_\eps
  - \dot{U}_\eps\,
  \big( H\, \delta_\eps\, U_\eps \dot{\big)}\Big)\
  \frac{Z_{4\eps}}{\sigma a^2-U_\eps^2H\delta_\eps}\,, \nonumber
\end{align}
(which at the moment we only need for $a_\eps$, $c_\eps$, $d_\eps$, $f_\eps$, 
$h_\eps$ vanishing) Proposition \ref{prop:exmat} gives solution candidates 
$U_\eps, Z_{p\eps}$ of the corresponding components of \eqref{eq:geos:eps} on 
$[\alpha_\eps,\alpha_\eps+\eta]$. Furthermore, we are able to solve the 
$V$-equation on $[\alpha_\eps,\alpha_\eps +\eta]$ since it is linear (using the 
already constructed solution candidates $U_\eps, Z_{p\eps}$). Moreover, 
$(V_\eps)_\eps$ is locally bounded uniformly in $\eps$, as can be seen as in 
\cite[Proposition 4.1, p.\ 12f.]{SSLP:16}. This gives a unique net of solutions
of \eqref{eq:geos:eps}  
\begin{equation}\label{eq-sol-cand}
  \gamma_\eps:= (U_\eps,V_\eps,Z_{p\eps}) \text{ defined on }
  [\alpha_\eps,\alpha_\eps+\eta], 
\end{equation}
with data constructed from the seed family with data \eqref{eq:data0}, see 
\eqref{eq:model-data}.

This net hence constitutes a solution candidate for the system  
\eqref{eq:geos:G} of geodesic equations in the generalized spacetime $(M,g)$ 
with data constructed from the seed data \eqref{eq:data0} and defined on the 
interval $J_\eps=[\alpha_\eps,\alpha_\eps +\eta]$. First, for the purpose of 
constructing a local solution candidate we assume the data \eqref{eq:data0}
to be constant in $\eps$ and to be given by \eqref{eq:data0-conv}. That is 
we assume the data to be given by the \emph{single} seed 
geodesic $\gamma^-$ with $\gamma^-(0)=(0,V^0,Z^0_p)$ and 
$\dot \gamma^-(0)= (\dot U^0,\dot V^0,\dot Z^0_p)$, which leads to the 
data 
\begin{equation}\label{eq:data-ro}
 \gamma_\eps(\alpha_\eps)=\gamma^-(\alpha_\eps),\quad
 \dot\gamma_\eps(\alpha_\eps)=\dot\gamma^-(\alpha_\eps)
\end{equation}
for the net $\gamma_\eps$ at $\alpha_\eps$. Next observe that the solution 
interval $J_\eps$ is depending on $\eps$ but in a favorable way. Indeed
we have by the definition \eqref{def-Xe} of $\Xe$  that 
\begin{equation}
U_\eps(\alpha_\eps+\eta) = -\eps + \int_{\alpha_\eps}^{\alpha_\eps+\eta} 
\dot{U}_\eps(s)\,\dd s
\geq -\eps + \frac{\eta}{2} \dot{U}^0\geq -\eps + 3\eps \geq  \eps\,,
\end{equation}
where in the one but last inequality we have used $\eps\leq\eta\,\dot 
U^0\!/6$ according to \eqref{eeq-eps-0}.

Summing up we have found a net of solutions $(\gamma_\eps)_\eps$ to 
\eqref{eq:geos:eps} with data \eqref{eq:data-ro} which (for $\eps$ small 
enough) crosses the regularization wave zone. More precisely call the parameter 
value when $\gamma_\eps$ leaves the wave zone $\beta_\eps$, i.e., 
$U_\eps(\beta_\eps)=\eps$, then $\gamma_\eps$ is defined on an interval that 
contains $[\alpha_\eps,\beta_\eps]$ in its interior since outside this interval 
$\gamma_\eps$ reaches the region of spacetime coinciding with the background 
(anti-)de Sitter space. 

To the 
past of $\alpha_\eps$ the net $\gamma_\eps$ smoothly extends to the seed 
geodesic $\gamma^-$ as long as the latter is defined. Being a background 
geodesic it will be past complete and it will also be a complete solution in 
the regularized impulsive wave spacetime as long as it does not reach the 
regularized wave zone (again). This will be for all negative values of 
the parameter in case $\sigma e\leq 0$ and at least for some finite 
negative parameter value in case $\sigma e>0$. In any case the net 
$\gamma_\eps$ extends smoothly to the single seed geodesic $\gamma^-$ in the 
region `before' the wave zone. 

To the future of $\beta_\eps$ the situation is slightly more complicated. 
Indeed $\gamma_\eps$ extends smoothly to a family of background geodesics 
$\gamma^+_\eps$ which are determined by the data of $\gamma_\eps$ at 
$\beta_\eps$, i.e., we have 
\begin{equation}\label{eq:data+}
\gamma^+_\eps(\beta_\eps)
 =\gamma_\eps(\beta_\eps)
 =:(\eps,V^{0+}_\eps,Z^{0+}_{p\eps } )\, ,
 \quad
 \dot \gamma^+_\eps(\beta_\eps)=\gamma_\eps(\beta_\eps)
  =:(\dot U_\eps^{0+},\dot V^{0+}_\eps,\dot Z^{0+}_{p\eps})\,.
\end{equation}
%
Now again in case $\sigma e\leq 0$ each $\gamma^+_\eps$ is (forward) complete 
and we have already found a \emph{global} solution candidate. However, in case 
$\sigma e>0$ we will again only have some finite parameter value before 
$\gamma^+_\eps$ reenters the wave zone. So in total we have at least found a 
local solution candidate that crosses the wave zone once and extends as a 
single seed geodesic into the background `before` the impulse and as a family 
of background geodesics into part of the (anti-)de Sitter spacetime `behind' 
the impulse. This situation is depicted in the left half of Figure 1
and we summarize it as follows:   

\begin{prop}[Extension of the solution candidate]\label{prop:global}
  The unique smooth geodesics $(\gamma_\eps)_\eps$ of \eqref{eq-sol-cand} 
  with initial data \eqref{eq:data-ro} extend to geodesics of the 
  background (anti-)de Sitter spacetime `before' and `behind' the wave 
  zone. Hence it provides a local solution candidate that crosses the 
  impulse once.
\end{prop}

\subsection{Construction of a global solution candidate}
Now we are going to globalize the above result, in the sense that we handle 
the case of infinitely many crossings of the wave impulse, 
thereby extending the maximal interval of definition of these solution 
candidates to all of $\R$. As discussed above this is only necessary if $\sigma 
e >0$, which will be assumed throughout this subsection. To this end we will 
employ advanced construction methods for (Colombeau) generalized functions. 
We make use of the following \emph{globalization Lemma}, c.f.\ \cite[Lemma 4.3, 
p.\ 12]{HKS:12} (see also \cite[Lemma 2.4, p.\ 6]{KSSV:14} but note that 
actually point (ii) there is redundant).

\begin{lem}\label{globallem} Let $M$, $N$ be manifolds, and $J:=(0,1)$. 
  Let $u\colon J\times M \to N$ be a smooth map and let (P) be a property 
attributable to values $u(\eps,p)$, satisfying:
For any $K\Subset M$ there exists some $\eps_K>0$ such that (P) holds for 
all $p\in K$ and $\eps<\eps_K$.
  Then there exists a smooth map $\tilde u\colon J\times M \to N$   such that 
(P) holds for all $\tilde u(\eps,p)$ ($\eps\in 
J$, $p\in M$) and for each $K\Subset M$ there exists some $\eps_K\in J$ such 
that $\tilde u(\eps,p) = u(\eps,p)$ for all 
$(\eps,p)\in (0,\eps_K] \times K$.
\end{lem}

\begin{thm}[Global solution candidate]\label{thm-glob-ex}
There exists a net of solutions $(\gamma_\eps)_\eps$ of \eqref{eq:geos:eps}, 
with data \eqref{eq:data-ro}, defined on all of $\R$ for all $\eps\in I$, 
thus constituting a global solution candidate for the system \eqref{eq:geos:G}.
\end{thm}

\begin{pr}
We will only prove forward completeness, i.e., existence on $[0,\infty)$, the 
case of backward completeness is completely analogous. We aim at applying Lemma 
\ref{globallem}. To this end we need to establish that we can apply Proposition 
\ref{prop:exmat} again after we already crossed the impulse once. That is we 
have to show that the geodesics $\gamma_\eps$ which `behind' the wave zone  
coincide with the background geodesics $\gamma^+_\eps$, see \eqref{eq:data+} 
qualify as a seed family (as defined in the beginning of Subsection 
\ref{subsec:sol-cand}). 

To formalize this denote the parameter value when $\gamma_\eps=\gamma^+_\eps$ 
(re)enters the regularization strip by $\alpha_\eps'$, 
i.e., $U_\eps(\alpha_\eps')=\eps$ (of course there is now a sign change). Then 
$\gamma_\eps=\gamma_\eps^+$  on $[\beta_\eps,\alpha_\eps']$ and we have to 
prove that $\gamma_\eps(\alpha_\eps')=\gamma^+_\eps(\alpha_\eps')$ 
and $\dot\gamma_\eps(\alpha_\eps')=\dot\gamma^+_\eps(\alpha_\eps')$ 
converge.
%

First observe that on $[\beta_\eps,\alpha_\eps']$ being a background geodesic 
we can explicitly calculate the $U$ component of $\gamma_\eps=\gamma^+_\eps$ to 
be (cf.\ e.g.\ \cite[Equation (3.2), p.\ 8]{SSLP:16})
\begin{equation}
 U_\eps(t) = a \dot{U}^{0+}_\eps \sin(\frac{t}{a})\,.
\end{equation}
So the next crossing of the impulse happens at $t=a\pi$ hence $\alpha_\eps'\to 
\pi a$, see Figure 1 for an illustration. 

Now by continuous dependence of solutions on the data,  on every compact 
subinterval $[t_1,t_2]$ of $[0,\pi a]$ the geodesics 
$\gamma_\eps=\gamma_\eps^+$ convergence uniformly together with their first 
derivatives to the background geodesic $\gamma^+$ with initial data at $t=0$ 
given by the limit of \eqref{eq:data+} (cf.\ \cite[Theorem 5.1, 
p.\ 15]{SSLP:16}). The latter indeed exists and has been explicitly related to 
the corresponding seed data in \cite[Proposition 5.3, p.\ 18]{SSLP:16}. 
Moreover this convergence is uniform with respect to $t_2$, i.e., we have for fixed 
$t_1$ and any $t_2\leq a\pi$
\begin{align}\nonumber\label{eq:max2}
 \sup_{t_1\leq t\leq t_2}\big(|\gamma_\eps(t)-\gamma^+(t)|&,
  |\dot \gamma_\eps(t)-\dot \gamma^+(t)|\big)
 \nonumber\\ 
 &\leq
 \max \big(|\gamma_\eps(t_1) - \gamma^+(t_1)|,
  |\dot{\gamma}_\ep(t_1) -
  \dot{\gamma}^+(t_1)|\big)\, e^{\pi a L}\ =: A(\eps)\, C
\end{align}
where $L$ is a Lipschitz constant for the right hand side of the geodesic 
equation of the background on a suitable compact set, $C$ is a constant and 
$A(\eps)\to 0$ for $\eps\to 0$. Hence we obtain
\begin{align}\nonumber
 \max&\big(|\gamma_\eps(\alpha'_\eps)-\gamma^+(a\pi)|,
 |\dot\gamma_\eps(\alpha'_\eps)-\dot\gamma^+(a\pi)|\big)\\ \nonumber 
 &\leq\max\big(
  |\gamma_\eps(\alpha'_\eps)-\gamma^+(\alpha'_\eps)|
  +|\gamma^+(\alpha'_\eps)-\gamma^+(a\pi)|,
  |\dot\gamma_\eps(\alpha'_\eps)-\dot\gamma^+(\alpha'_\eps)|
  +|\dot\gamma^+(\alpha'_\eps)-\dot\gamma^+(a\pi)|\big)
  \\\nonumber
 &\leq A(\eps)\,C\,\max\big(|\gamma^+(\alpha'_\eps)-\gamma^+(a\pi)|,
  |\dot\gamma^+(\alpha'_\eps)-\dot\gamma^+(a\pi)|\big)\,.
\end{align}
Now the latter term converges to zero by smoothness of $\gamma^+$ and we have 
established that $\gamma^+_\eps(\alpha_\eps')=\gamma_\eps(\alpha_\eps') \to 
\gamma^+(\pi a)$ and 
$\dot\gamma^+_\eps(\alpha_\eps')=\dot{\gamma}_\eps(\alpha_\eps') \to 
\dot{\gamma}^+(\pi a)$. In conclusion, the net $(\gamma^+_\eps)_\eps$ is a seed 
family and we can (re)apply the machinery of the previous subsection, in 
particular, Proposition \ref{prop:exmat}. However, the corresponding $\eps_0$ 
now depends on (the limit of) \eqref{eq:data+}. 
\medskip

To iterate this construction set $K_n:=[-\frac{a\pi}{2},n \frac{a\pi}{2}]$ for 
$1\leq n\in\N$ which gives a compact exhaustion of 
$[-\frac{a\pi}{2},\infty)$ (i.e., $K_n\subseteq K_{n+1}^\circ$ and 
$\bigcup_{n\in\N}K_n=[-\frac{a\pi}{2},\infty)$), such that $K_n$ yields exactly 
$n$ crossings ($n \geq 1$). Then by the above for every $n\in\N$ there is an 
$\eps_n>0$ such that for every $0<\eps\leq\eps_n$ there is a unique solution 
$\gamma_\eps$ of \eqref{eq:geos:eps} with initial data \eqref{eq:data-ro} 
on $K_n$. 
\medskip

At this point we apply Lemma \ref{globallem} by setting
$M:=[-\frac{a\pi}{2},\infty)$ and $N:=\R^5$ and define
$\gamma\colon (0,\infty)\times [-\frac{a\pi}{2},\infty)\to \R^5$ by
\begin{equation} 
(\eps,t)\mapsto 
\gamma_\eps(t)\,,\ \text{whenever}\ t\in K_n
 \text{, where }n\text{ is minimal and }0<\eps\leq\eps_n
\end{equation}
and extend it arbitrary but smoothly to bigger values of $t$ and $\eps$. 
Then from the above (in particular, the uniqueness of the solutions 
of \eqref{eq:geos:eps}, \eqref{eq:data-ro}) it is clear that if $\gamma_\eps$ 
is a solution on $K_n$, then $\gamma_\eps$ is also a solution on $K_m$ for 
$m\leq n$. At this point we define the property (P) by 
\begin{equation*}
 \text{(P) holds at } \gamma_\eps(t) \text{ if } 
  \gamma_\eps(t) \text{ solves \eqref{eq:geos:eps} at } t\,.
\end{equation*}

Now the assumptions of the lemma hold since for any $K\Subset [0,\infty)$ 
there is a minimal $n\in\N$ such that $K\subseteq K_n$. Thus there 
is an $\eps_n>0$ and there are unique solutions $(\gamma_\eps)_{\eps\leq\eps_n}$ 
on $K_n$, hence on $K$. So 
Lemma \ref{globallem} provides $\tilde{\gamma}_\eps$ (again called 
$\gamma_\eps$ in the statement of the theorem) defined on $[0,\infty)$ that is a 
solution of the geodesic equations \eqref{eq:geos:eps} with data 
\eqref{eq:data-ro} for all $\eps$.
%
%
\end{pr}

\begin{figure}\label{fig:reg}
 \includegraphics{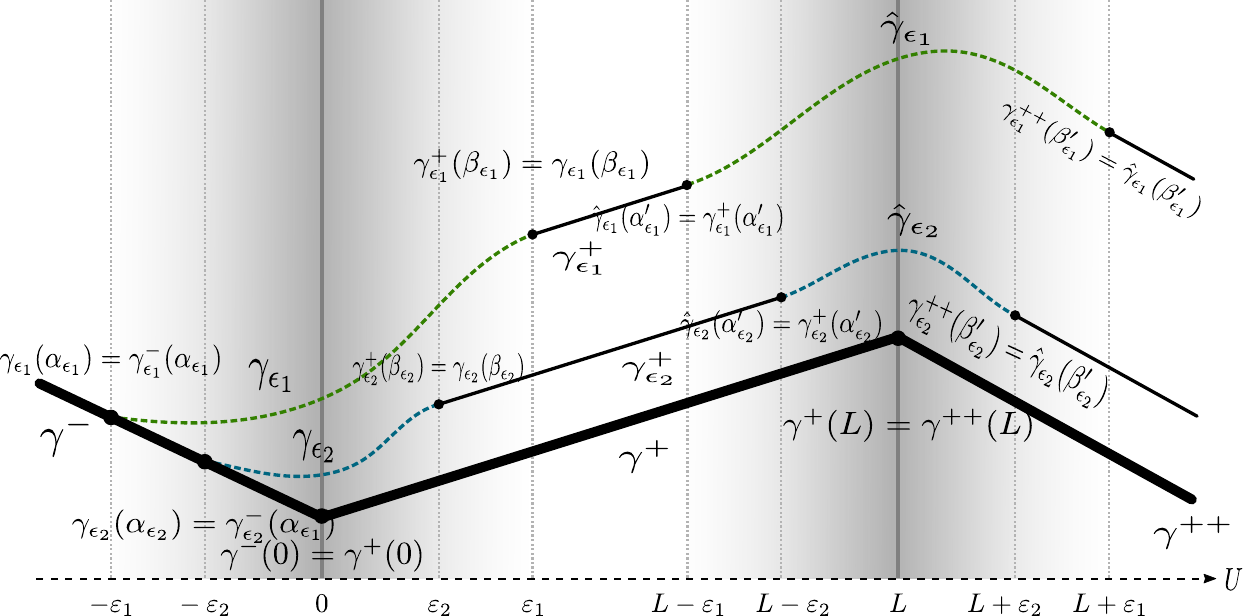}
 \caption{The construction of the solution candidate for the first two 
crossings of the impulse at parameter values $t=0$ and $t=L=a\pi$.}
\end{figure}

\subsection{Global existence of solutions}

In this section we establish existence of solutions of the geodesic equations 
\eqref{eq:geos:G}. Indeed, by showing c-boundedness and moderateness of the 
solution candidate given in Theorem \eqref{thm-glob-ex} we actually prove 
existence of \emph{global} solutions.
\begin{thm}[Global Existence]\label{prop-ex-G}
 The global solution candidate $(\gamma_\eps)_{\eps\in I}$ of \eqref{eq:geos:G}, 
 given by Theorem \ref{thm-glob-ex} is moderate and c-bounded. Hence it defines
 a global solution $[(\gamma_\eps)_\eps]\in\G[\R,M]$ to the geodesic 
 equation \eqref{eq:geos:G} with data \eqref{eq:data-ro} (derived from a single
 seed geodesic with data \eqref{eq:data0-conv}). 
\end{thm}

\begin{pr}
First observe that we have to prove the asymptotic estimates on compact time 
intervals only. So we can restrict ourselves to such intervals that contain one 
single crossing of the impulse or to such which contain no crossing at all.

To deal with the first case without loss of generality we only consider the 
first crossing of the impulse at $t=0$. Let $(\gamma_\eps)_{\eps}$ be given by 
Theorem \ref{thm-glob-ex}, then $\gamma_\eps$ is a solution of 
\eqref{eq:geos:eps} on $[\alpha_\eps,\alpha_\eps+\eta]$, where $\eta$ is 
independent of $\eps$. On this interval the $U$- and $Z$-components (i.e., 
$X_\eps=(U_\eps,Z_\eps)$) and their first order derivatives are are bounded 
uniformly in $\eps$ by Proposition \ref{prop:exmat}. So they are c-bounded 
together with their first order derivatives. Moreover by iteratively using the 
differential equation \eqref{eq:geos:eps} (with coefficients at worst of order 
$1/\eps$) the derivatives of order $k$ satisfy an $O(\eps^{1-k})$ estimate. 
Also, $V_\eps$ is bounded, uniformly in $\eps$, on the interval 
$[\alpha_\eps,\beta_\eps]$ (cf.\ the discussion above equation 
\eqref{eq-sol-cand}) and so by integration, $\dot{V}_\eps$ satisfies an 
$O(\eps^{-1})$ estimate. Inductively, the higher order estimates again follow 
from \eqref{eq:geos:eps}.

Now let $K$ be a compact time interval disjoint from any crossing of 
the impulse, i.e., $ka\pi\not\in K$ for all $k\in\Z$. There $(\gamma_\eps)_\eps$ are 
background geodesics depending on $\eps$ only via their data. 
Since the latter converges we obtain moderateness and c-boundedness between 
impulses as follows. Via a mean value argument one sees that $\gamma_\eps$ and 
$\dot{\gamma}_\eps$ are uniformly bounded (cf.\ \cite[Equation (3.10), p.\ 
9]{SSLP:16}). Differentiating the differential equation \eqref{eq:geos:eps} 
${k-2}$-times and using again continuous dependence on initial conditions yields 
that the $k$-th derivative satisfies an $O(\eps^{-k+1})$ estimate. This 
establishes moderateness, c-boundedness is clear by the uniform 
boundedness properties of $\gamma_\eps$ as noted above.
\end{pr}

\subsection{Global uniqueness of solutions}\label{subsec:uniq}

Our final step is to show that the global geodesics obtained above are in fact  
the \emph{unique} solutions of \eqref{eq:geos:G} with corresponding data 
\eqref{eq:data-ro}. As remarked in the beginning of this section this  can be 
viewed as an additional stability result.

First note that uniqueness for \eqref{eq:geos:G}, \eqref{eq:data-ro} does not 
follow from the uniqueness part of Proposition \ref{prop:exmat} since 
this only provides unique solvability of \eqref{eq-ivp-eps} respectively  
\eqref{eq:geos:eps}, \eqref{eq:data-ro}, that is uniqueness 'on the 
$\eps$-level'. Here we have, however, to provide uniqueness of 
\eqref{eq:geos:G}, \eqref{eq:data-ro} in $\G[\R,M]$, that is we have to 
show that negligibly perturbed data and right-hand-side of the equations only lead to 
negligibly perturbed solutions.

We will even show uniqueness in $\G[\R,\R^5]$ by using the result for the 
(perturbed) model IVP \eqref{eq-ivp-eps}. In fact, this is stronger than needed, 
since uniqueness in $\G[\R,M]$ would restrict the possible negligible 
perturbations (so as to stay on the hyperboloid $M$). 

Moreover note that for uniqueness in $\G[J,\R^5]$ we would need also to perturb 
the $U$-initial data, i.e., $u_\eps(\alpha_\eps)=-\eps + n_\eps$, where 
$(n_\eps)_\eps\in \NN$. However, by shifting the parameter (e.g., 
$u_\eps(\alpha_\eps + A_\eps) = -\eps$) we can without loss of generality assume 
that there is no negligible perturbation in the $U$-initial data. 

\begin{thm}[Uniqueness] 
The global solution $[(\gamma_\eps)_\eps]\in\G[\R,M]$ 
of Theorem \ref{prop-ex-G} to the geodesic equation \eqref{eq:geos:G} with 
data \eqref{eq:data-ro} is unique. 
\end{thm}

\begin{pr}   
To begin with we observe that it is sufficient to consider the model system 
\eqref{eq-ivp-eps} since the $V$-equation can simply be integrated once $U$ and 
$Z_p$ are known. This gives uniqueness by the basic fact that integration is 
well defined in generalized functions.

So let $x_\eps$ be a solution of the initial value problem \eqref{eq-ivp-eps} 
with all negligible terms set to zero and 
let $\tilde{x}_\eps:=(\tilde{u}_\eps,\tilde{z}_\eps)$ solve 
\eqref{eq-ivp-eps} with all the negligible terms on the r.h.s.\ and the data in 
effect. Generally we will denote quantities depending on $\tilde{x}_\eps$ also 
with a tilde.

Again we have to show the asymptotic estimates only on compact time intervals 
$J$ containing at most one crossing of the impulse. Without loss of generality 
we deal with the first crossing at $t=0$. Let  $q\in\N$.
\medskip

We start with the $U$-component: Since $a$ and $d$ are 
negligible, there is $C>0$ and such that for all 
$\eps$ small enough and for all $t \in J$
\begin{align}\label{eq:uunique}
  |u_\eps(t)-\tilde{u}_\eps(t)|&\leq C\eps^q + 
\int_{\alpha_\eps}^t\int_{\alpha_\eps}^s \bigl|\frac{u_\eps}{N_\eps} - 
\frac{\tilde{u}_\eps}{\tilde{N}_\eps}\bigr|\dd r \dd s+ 
\int_{\alpha_\eps}^t\int_{\alpha_\eps}^s \bigl|\frac{\Delta_\eps 
u_\eps}{N_\eps} - 
\frac{\tilde{\Delta}_\eps\tilde{u}_\eps}{\tilde{N}_\eps}\bigr|\dd r \dd s\,.
\end{align}
Using Lemma \ref{lem-1.-est}(i) the first integrand can be bounded by 
$\frac{4}{a^4}|u_\eps| |\tilde{u}_\eps^2 \tilde{H}\tilde\delta_\eps - u_\eps^2 H 
\delta_\eps|+\frac{2}{a^2}|u_\eps-\tilde u_\eps|$. Keeping the second term and 
estimating the first one further we obtain
\begin{align}\label{eq:Nest}
 |\tilde{u}_\eps^2 \tilde{H}\tilde\delta_\eps - u_\eps^2 H \delta_\eps| \leq |u_\eps - \tilde{u}_\eps|
|\tilde{u}_\eps\tilde{H}\tilde{\delta}_\eps| + |\tilde{u}_\eps u_\eps \tilde{\delta}_\eps| |H-\tilde{H}| + |H u_\eps| 
|\tilde{u}_\eps\tilde{\delta}_\eps - u_\eps \delta_\eps|\,,
\end{align}
where we can use $|H-\tilde{H}|\leq \lip(H)|z_\eps-\tilde{z}_\eps|$ for the second term.

For the third term we use a generalization of \cite[Lemma 
A.4(i), p.\ 25]{SSLP:16}. Note that here we have 
$x_\eps\in\Xe\not=\tilde{\Xe}\ni \tilde{x}_\eps$ since they have different 
initial conditions. 
Nevertheless the estimate in Lemma \ref{lem-1.-est}(ii) on the 
diameter of $\Gamma_\eps(u_\eps)$ is independent of $d_\eps$ and so
\begin{align*}
\diam(\Gamma_\eps(u_\eps)\cup\Gamma_\eps(\tilde{u}_\eps))\leq 
\frac{4\eps}{\dot{u}^0}
\end{align*}
and the proof of \cite[Lemma A.4(i), p.\ 25]{SSLP:16} remains intact. Hence we 
obtain
\begin{align*}
  \int_{\alpha_\eps}^t |\tilde{u}_\eps\tilde{\delta}_\eps - u_\eps 
\delta_\eps| \dd s   \leq C |\tilde{u}_\eps(t)-u_\eps(t)|\,.
\end{align*}
Summing up we have
\begin{align}
  \int_{\alpha_\eps}^t\int_{\alpha_\eps}^s \bigl|\frac{u_\eps}{N_\eps} - 
  \frac{\tilde{u}_\eps}{\tilde{N}_\eps}\bigr|\dd r \dd s
  \leq C \psi(t)\,,
\end{align}
where $\psi(t):= |u_\eps(t) - \tilde{u}_\eps(t)| + |\dot{u}_\eps(t) - 
\dot{\tilde{u}}_\eps(t)| + |z_\eps(t) - 
\tilde{z}_\eps(t)| + |\dot{z}_\eps(t) - \dot{\tilde{z}}_\eps(t)|$. 

The second integrand of \eqref{eq:uunique}  can be controlled as follows
\begin{align*}
 \bigl|\frac{u_\eps\Delta_\eps}{N_\eps} - \frac{\tilde{u}_\eps \tilde{\Delta}_\eps}{\tilde{N}_\eps}\bigr|\leq 
\frac{2}{a^2}|u_\eps \Delta_\eps - \tilde{u}_\eps\tilde{\Delta}_\eps| + 
\frac{4}{a^4}|\tilde{u}_\eps\tilde{\Delta}_\eps||\tilde{N}_\eps - N_\eps|\,,
\end{align*}
where the first term on the right-hand-side can be bounded as in 
\cite[(A.35), p.\ 27]{SSLP:16} using the above generalization of \cite[Lemma A.4(i), p.\ 
25]{SSLP:16}. The second term is proportional to the one estimated in 
\eqref{eq:Nest} and we obtain
\begin{align*}
|u_\eps(t)-\tilde{u}_\eps(t)|&\leq C \eps^q + C 
\int_{\alpha_\eps}^t\int_{\alpha_\eps}^s \psi(r) \dd r \dd s\,.
\end{align*}

The estimates for $|\dot u_\eps(t)-\dot{\tilde{u}}_\eps(t)|$ are 
obtained in complete analogy and give
\begin{align*}
 |\dot{u}_\eps(t) - \dot{\tilde{u}}_\eps(t)|&\leq C \eps^q + C \int_{\alpha_\eps}^t 
\psi(s) \dd s\,.
\end{align*}
\medskip

The estimates for $|z_\eps(t) - \tilde{z}_\eps(t)|$ and its derivative 
are obtained along the same lines now using an analogous 
generalization of \cite[Lemma A.4(ii), p.\ 25]{SSLP:16} and give
\begin{align*}
 |z_\eps(t) - \tilde{z}_\eps(t)|&\leq C \eps^q + C \int_{\alpha_\eps}^t \int_{\alpha_\eps}^s \psi(r) \dd r \dd s\,,\\
 |\dot{z}_\eps(t) - \dot{\tilde{z}}_\eps(t)|&\leq C \eps^q + C \int_{\alpha_\eps}^t \psi(s) \dd s + C \int_{\alpha_\eps}^t 
\int_{\alpha_\eps}^s \psi(r) \dd r \dd s +  C \int_{\alpha_\eps}^t \int_{\alpha_\eps}^s \int_{\alpha_\eps}^r \psi(\tau) \dd 
\tau \dd r \dd s\,.
\end{align*}
Here we have used that $x_\eps$ and $\tilde{x}_\eps$ are the fixed points of 
the corresponding operators \eqref{eq-A_eps} $A_\eps$ and $\tilde{A}_\eps$, respectively. Thus $z_\eps = 
A_\eps^2(A_\eps(x_\eps))$ and $\tilde{z}_\eps = \tilde{A}_\eps^2(\tilde{A}_\eps(\tilde{x}_\eps))$ and we can eliminate any 
$\frac{1}{\eps}$-terms, in contrast to \cite[Proposition A.5, 
p.\ 26ff.]{SSLP:16}. 
\medskip

So finally,
\begin{align*}
 \psi(t)\leq C \eps^q + C \int_{\alpha_\eps}^t \psi(s) \dd s + C \int_{\alpha_\eps}^t 
\int_{\alpha_\eps}^s \psi(r) \dd r \dd s +  C \int_{\alpha_\eps}^t \int_{\alpha_\eps}^s \int_{\alpha_\eps}^r \psi(\tau) \dd 
\tau \dd r \dd s\,,
\end{align*}
and consequently $\sup_{t\in J}\psi(t)\leq C\eps^q$ by Bykov's 
inequality (a generalization of Gronwall's inequality, cf.\ \cite[Theorem 
1.11, p.\ 98f.]{BS:92}). Using \cite[Lemma 1.2.3, p.\ 11]{GKOS:01} we obtain 
that $(u_\eps-\tilde{u}_\eps)_\eps$ and $(z_\eps - \tilde{z}_\eps)_\eps$ are 
negligible.
\end{pr}

\section{Geodesic completeness}\label{sec:compl1}

We now address the issue of completeness and begin with a discussion of the 
geodesics neglected so far, i.e., those whose initial data is \emph{not} given 
by a seed geodesic crossing $\{U=0\}$. These remaining geodesics 
$\gamma=(U,V,Z_p)$ all have a trivial $U$-component hence propagate parallel to 
the impulsive wave located at $\{U=0\}$. In case $U=\text{const.}\not=0$ 
such geodesics will never enter the regularized wave zone for $\eps$ 
small enough. Hence they are background geodesics and therefore complete. 
Finally to treat geodesics with $U\equiv 0$, observe that the surface $\{U=0\}$ 
is 
totally geodesic (not only in the background but also) in the generalized 
spacetime, which can be seen from the $U$-component of the geodesic equations 
\eqref{eq:geos:G} (see also \cite{PSSS:15}, the discussion prior to Theorem 3.6). 
Hence such geodesics have trivial $U$-components and consequently  the system 
\eqref{eq:geos:G} reduces to the background geodesic equations which again 
leads to completeness.

So the only real issue are the geodesics crossing the shock which we have 
already dealt with in Section \ref{sec:complete} above. Recall that there we did not put any restrictions on the cosmological 
constant $\Lambda$ or the causal character $e$ of the geodesics. So we obtain without 
further effort our main result, the geodesic completeness  of all non-expanding 
impulsive gravitational waves propagating in the (anti-)de Sitter universe. 

\begin{cor}[Completeness]
 The generalized impulsive wave spacetime $(M,g)$ given by \eqref{eq:M}, 
\eqref{eq:Mbar}  is geodesically complete.  
\end{cor}

This result is the desired generalization of the completeness result for impulsive \emph{pp}-waves in \cite{KS:99} to the 
case $\Lambda\not=0$. Also it is related to the completeness result \cite[Thm.\ 5.3]{SSS:16}, where the latter is formulated 
in the spirit of \cite{SSLP:16} without the use of nonlinear generalized functions. This result applies to a more general 
class of impulsive waves including gyratonic terms and allowing for a non-flat wave surface, but necessarily have vanishing 
$\Lambda$. Observe, however, that the techniques developed in the present paper 
rely heavily on the $5$-dimensional embedding formalism and hence differ 
significantly from the ones used in the ${\Lambda=0}$-cases 
of \cite{KS:99,SSLP:16}.

\section{Associated geodesics}\label{sec:4}

In this final section we briefly demonstrate how we can recover the limiting 
geodesics of \cite[Section 5]{SSLP:16} in our present framework. We will do so 
using the notion of association in spaces of generalized functions briefly 
recalled in Section \ref{sec:col}. We start by making the latter more precise 
and additionally define the notion of $k$-association.

\begin{defi}
 Let $\Omega\subseteq \R^n$ be open. A generalized function 
$f=[(f_\eps)_\eps]\in\G(\Omega)$ is called 
\begin{enumerate}
 \item[(i)] \emph{associated with $u\in\D'(\Omega)$}, denoted by $f\approx u$, 
if
\begin{equation*}
 \lim_{\eps\to0} \int_\Omega f_\eps(x)\phi(x)\,\dd x = \lara{u,\phi}\quad
 \text{for all $\phi\in\D(\Omega)$}\,,
\end{equation*}
where $\langle\ ,\ \rangle$ denotes the distributional action.
\item[(ii)] \emph{$k$-associated with $u\in C^k(\Omega)$}, denoted by 
$f\approx_k u$, if $f_\eps\to u$ in $C^k(\Omega)$, i.e., uniformly on all 
compact subsets of $\Omega$ up to derivatives of order $k$. 
\end{enumerate}
\end{defi}

Next we define the limiting geodesics which can heuristically be viewed as the 
geodesics of the distributional form of the metric given by \eqref{classical}, 
\eqref{const}. Note, however, that these do \emph{not} actually solve the 
corresponding geodesic equations. These limiting geodesics are given by 
appropriately matched geodesics of the background spacetime across the 
impulsive wave surface. For simplicity we only deal with the first crossing of 
the impulse at $t=0$. Let  
$[(\gamma_\eps)_\eps]=[(U_\eps.V_\eps,Z_{p\eps})_\eps]$ be the 
(global) solution of \eqref{eq:geos:eps}, \eqref{eq:data-ro}, given by Theorem 
\ref{thm-glob-ex} and set  
\begin{equation}\label{eq:limgeos}
 \tilde\gamma(t)=(\tilde U,\tilde V,\tilde Z_p)(t):=\begin{cases}
  \gamma^-(t)\,,\qquad&t \leq 0\\
  \gamma^+(t)\,,\qquad&t>0\,,\end{cases}
\end{equation}
where $\gamma^-=(U^-,V^-,Z^-_{p})$ is the seed geodesic with data 
\eqref{eq:data0-conv}, that is  
\begin{equation}\label{eq:sd}
 \gamma^-(0)=(0,V^0,Z^0_p)\,,\quad\text{and}\quad 
 \dot \gamma^-(0)= (\dot U^0>0,\dot V^0,\dot Z^0_p)\,,
\end{equation}
(cf.\  the discussion prior to \eqref{eq:data-ro}). 
Furthermore $\gamma^+=(U^+,V^+,Z^+_{p})$ is 
the background geodesic crossing $U=0$ at $t=0$ with data 
given by the limit of \eqref{eq:data+}, that is more precisely
\begin{equation}\label{eq:Xdata_ro}
\gamma^+(0)=(0,{\sf B},Z_p^0)\,,\quad \text{and}\quad
\dot \gamma^+(0)=(\dot U^0,{\sf C},{\sf A_p})\,,
\end{equation}
where 
\begin{equation}\label{eq:limconst}
 {\sf A_p}:=\lim_{\eps\to 0} \dot Z_{p\eps}(\beta_\eps)\,,\quad
    {\sf B}=\lim_{\eps\to 0} V_\eps(\beta_\eps)\,,\quad
    {\sf C}=\lim_{\eps\to 0} \dot V_\eps(\beta_\eps)\,.
\end{equation}
Again $\beta_\eps$ 
is defined to be the time when $\gamma_\eps$ leaves the regularization strip, 
i.e., $U_\eps(\beta_\eps)=\eps$. We also remark that the constants ${\sf A_p}$, 
${\sf B}$, and ${\sf C}$ have been explicitly computed in terms of the seed 
data \eqref{eq:sd} and the values of $H$ and its first order derivatives at the 
impulsive surface in \cite[Proposition 5.3]{SSLP:16}.

Now \cite[Theorem 5.1, p.\ 14ff.]{SSLP:16} translates into the following 
statement.

\begin{thm}[Associated geodesics]
The solution $\gamma=(U,V,Z_p)=[(\gamma_\eps)_\eps]$ of \eqref{eq:geos:G}, 
\eqref{eq:data-ro} is associated with the limiting geodesic 
$\tilde \gamma$ of \eqref{eq:limgeos}. Moreover we have $U\approx_1\tilde U$ 
and $Z_p\approx_0\tilde Z_p$. 
\end{thm}

Note that these convergences are optimal in the light of $\tilde V$ and 
$\dot{\tilde{Z}}_p$ being discontinuous across $t=0$, i.e., the limiting 
geodesics being refracted geodesics of the background suffering a jump in the 
$V$-position and $V$-velocity as well as in the $Z_p$-velocity, cf.\ 
\cite[Section 5]{SSLP:16}.  
\medskip

Finally we discuss our completeness result in the light of the standard result 
(e.g.\ \cite[Rem.\ 3.5]{CS:08}) in smooth semi-Riemannian geometry which says 
that for an incomplete geodesic $\lambda$ necessarily $\lambda'$ leaves any 
compact subset of the tangent bundle. On the other hand even if $\lambda$ is 
complete and stays in a compact set in the manifold, $\lambda '$ need not be 
contained in a compact subset of the tangent bundle. The situation in the 
generalized setting is more subtle as can be seen from the spacetime at hand.

First note that while in the background (anti-)de Sitter spacetime the 
spacelike (respectively timelike) geodesics are closed this is \emph{not} the 
case in the impulsive spacetime due to the jump of the limiting geodesic 
$\tilde \gamma$ in the $V$-position on crossing the impulse. Moreover the jump
will in general depend on the specific profile function $H$ and $\tilde \gamma$ 
will in general \emph{not} stay in a compact set $K\Subset M$. 
Consequently, one cannot guarantee that the points $\gamma(t_n)$ when the 
generalized geodesic $\gamma=[(\gamma_\eps)_\eps]$ hits the impulse are 
confined to a compact subset of the hypersurface on which the wave is 
supported. 

Now we concentrate our attention to just one crossing of the impulse by 
a generalized geodesic $\gamma$. By the boundedness properties
established in \cite[Prop.\ 4.1(i)--(iii)]{SSLP:16} $\gamma_\eps$ stays in 
a compact subset of $M$ uniformly in $\eps$. However, one sees that by the jump 
in $V$ the boundedness results established in \cite[Prop.\ 
4.1(iv)]{SSLP:16} on $\dot V_\eps$ are optimal in the sense that one can only 
bound $\dot V_\eps(\beta_\eps)$, i.e., $\dot V_\eps$ at the instant it exits 
the wave zone but not throughout the entire wave zone. Hence $\dot \gamma_\eps$ 
will not be confined to a compact set of $TM$ uniformly in $\eps$ even on 
\emph{one single} crossing. So one can rather think of the situation already 
locally of being similar to the above, where one has complete geodesics in a 
compact set with their tangents \emph{not} contained in a compact set.

\section*{Acknowledgment}
We thank Ji\v{r}\'i Podolsk\'y for generously sharing his 
expertise. This work was supported by the Austrian Science Fund FWF, projects
P26859 and P28770. Also the authors acknowledge the hospitality of ESI during 
the programmes Geometry and Relativity as well as Geometric Transport Equations 
in General Relativity.

\end{document}